\documentclass[conference]{IEEEtran}
\IEEEoverridecommandlockouts

\usepackage[noadjust]{cite}

\usepackage[caption=false,font=footnotesize]{subfig}
\usepackage{graphicx}
\graphicspath{{./figure/}}

\usepackage{bm}
\usepackage{amsmath}
\usepackage{amsfonts}
\usepackage{amsmath}
\usepackage{amssymb}

\usepackage{lipsum}

\usepackage{algorithmic}

\usepackage{array}
\usepackage{multirow}
\usepackage{booktabs}

\usepackage[]{xcolor}

\usepackage{url}

\DeclareMathOperator*{\diag}{diag}
\DeclareMathOperator*{\argmin}{arg\,min}

\newcommand{\R}{\mathbb{R}}

\newcommand{\G}{\mathcal{G}}
\newcommand{\V}{\mathcal{V}}
\newcommand{\E}{\mathcal{E}}

\newcommand{\U}{\mathcal{U}}

\newcommand{\T}{\mathcal{A}}

\title{Infinite Impulse Response Graph Neural Networks for Cyberattack Localization in Smart Grids\\
	\thanks{This work was supported by NSF under Award Number 1808064.}
}

\begin{document}

\author{
	\IEEEauthorblockN{Osman Boyaci}
	\IEEEauthorblockA{
		Electrical Engineering\\
		Texas A\&M University\\
		College Station, TX, 77843\\
		osman.boyaci@tamu.edu
	}  \and
	\IEEEauthorblockN{M. Rasoul Narimani}
	\IEEEauthorblockA{
		College of Engineering\\
		Arkansas State University\\
		Jonesboro, AR, 72404\\
		mnarimani@astate.edu
	} \and
	\IEEEauthorblockN{Katherine Davis}
	\IEEEauthorblockA{
		Electrical Engineering\\
		Texas A\&M University\\
		College Station, TX, 77843\\
		katedavis@tamu.edu
	} \and
	\IEEEauthorblockN{Erchin Serpedin}
	\IEEEauthorblockA{
		Electrical Engineering\\
		Texas A\&M University\\
		College Station, TX, 77843\\
		eserpedin@tamu.edu
	}
}


\maketitle


\begin{abstract}
This study employs Infinite Impulse Response (IIR) Graph Neural Networks (GNN) to efficiently model the inherent graph network structure of the smart grid data to address the cyberattack localization problem. 
First, we numerically analyze the empirical frequency response of the Finite Impulse Response (FIR) and IIR graph filters (GFs) to approximate an ideal spectral response.
We show that, for the same filter order, IIR GFs provide a better approximation to the desired spectral response  and they also present the same level of approximation to a lower order GF due to their rational type filter response.
Second, we propose an IIR GNN model to efficiently predict the presence of cyberattacks at the bus level.
Finally, we evaluate the model under various cyberattacks at  both sample-wise (SW) and bus-wise (BW) level, and compare the results with the existing architectures. 
It is experimentally verified that the proposed model outperforms the state-of-the-art FIR GNN model by $9.2\%$ and $14\%$ in terms of SW and BW localization, respectively.
\end{abstract}


\section{Introduction}
Graph structural data such as electric grid networks, social networks, sensor networks, and transportation networks cannot be modeled efficiently in the Euclidean space and require graph-type architectures due to their inherent graph-based topologies \cite{wu2020comprehensive}.
Units are ordered and present the same number of neighbors in image/video data, therefore, they can be processed in an Euclidean space.
For instance, a sliding kernel can easily capture the spatial correlations of pixels in the Euclidean space.
In contrast, neighborhood relationships are unordered and vary from node to node in a graph signal \cite{wu2020comprehensive}.
Thus, graph signals need to be processed in non-Euclidean spaces determined by the underlying graph topology.
As a highly complex graph structural data, smart grid signals require graph architectures such as Graph Signal Processing (GSP) or Graph Neural Network (GNN) to exploit the spatial correlations.

To deal with the graph structural data in non-Euclidean spaces, GSP has emerged in the past few years \cite{ortega2018graph}.
In GSP, similar to the classical signal processing, a graph signal is first transformed into the spectral domain by Graph Fourier Transform (GFT), then its Fourier coefficients are scaled in the spectral domain, and finally the signal transformed back into the vertex domain by the inverse GFT \cite{shuman2013emerging}.
To circumvent the computationally complex domain transformation operations, Finite Impulse Response (FIR) graph filters (GFs) are proposed in \cite{defferrard2016convolutional} in which localized filters are learned directly in the vertex domain without any GFT operations \cite{bianchi2021graph}.
Spectral response of an FIR GF is a $K$-order polynomial since the output of each vertex $v$ is only dependent on the $K$-hop neighborhood of $v$.
Yet, to capture the global structure of a graph, FIR GFs may require  high degree polynomials since their frequency responses are not ``flexible'' enough to adapt to sudden changes in the spectral domain \cite{bianchi2021graph}. 
Nevertheless, the interpolation and extrapolation performance of high degree polynomials are not satisfactory \cite{shi2015infinite}. 
Infinite Impulse Response (IIR) GFs are proposed in \cite{shi2015infinite} to overcome the limitation of FIR GFs. 
Contrary to FIR GFs, IIR GFs present rational type spectral responses. 
Thus, IIR GFs can implement more complex responses with low-degree polynomials in the numerator and denominator because rational functions are more flexible than polynomial functions in terms of interpolation and extrapolation capabilities \cite{shi2015infinite, bianchi2021graph}. 

Information and Communication Technologies (ICT) are integrated into large-scale power networks to increase the efficiency of generation, transmission, and distribution systems~\cite{yu2016smart}.
Remote Terminal Units (RTUs) placed in electric power grids acquire the physical measurements and deliver them to the Supervisory Control and Data Acquisition Systems (SCADAs) and the ICT network transfers these measurements to the application level where the power system operators process them.
Consequently, the power system's reliability strongly depends on the accuracy of these steps along this cyber-physical pipeline.
Power system state estimation (PSSE) modules employ these measurements to estimate the current operating point of the grid~\cite{abur2004power}.
Besides, the accuracy of power system analysis tools such as energy management, contingency and reliability analysis, load and price forecasting, and economic dispatch depend on these measurements.
As a direct consequence, metering devices represent highly attractive targets for adversaries that try to obstruct the grid operation by corrupting the measurements.

False data injection attacks (FDIAs) constitute a significant portion of the cyber-physical threats to smart grids. FDIAs assumes inoculation of false data to the measurements to mislead the PSSE process.
Any action taken by the grid operator based on the false operating point can lead to serious consequences including systematic failures \cite{musleh2019survey}.

Numerous methods have been proposed to detect the presence of FDIA cyberattacks without providing information about their location \cite{musleh2019survey}. 
Cyberattack localization is critical for reliable grid operation and control since preventive actions including isolating the under-attack buses and re-dispatching the system can be taken.
For this reason, this paper focuses on cyberattack localization in smart grids.

The current approaches in cyberattack localization in power grids suffer from some limitations because it is a relatively new research topic compared to the detection of these attacks.
A multistage localization algorithm based on graph theory results is proposed in \cite{nudell2015real} to localize the attack at the cluster level.
Yet, cluster level algorithms limit the benefits of localization due to their low resolution. 
A model-driven analytical redundancy approach utilizing Kalman filters is presented in \cite{khalaf2018joint} for joint detection and mitigation of cyberattacks in Automatic Gain Control (AGC) systems. 
Authors in \cite{khalaf2018joint} determine a Mahalanobis norm based threshold of the residuals for the non-attacked case and residues larger than this threshold are regarded as attacked samples.
Apart from the manual threshold optimization steps, their detection times are in the range of seconds in their estimation based models. 	
A GSP based approach is developed in \cite{hasnat2020detection} to detect and localize the cyberattacks. 
Nevertheless, the random and easily detectable attack methodologies employed to test their models do not comprehensively assess the actual performance of the models.
Authors in \cite{jevtic2018physics} propose physics- and learning-based approaches to detect and localize cyberattacks in power systems.
They utilize a Long Short Term Memory (LSTM) Neural Network (NN) to generate a model for learning the data pattern.
Nonetheless, their results are limited to a 5-bus system and they train an LSTM model for each measurement.
The limited number of components deeply restrains the large-scale attributes of their proposed method since training a separate detector for each bus extremely increases the overall model complexity for large systems and reduces its suitability for real world applications.

Cyberattack localization can be a challenging task if an adversary has `enough' information about the grid \cite{liu2011false}.
Besides, if the designed GFs do not satisfy the required spectral response, the attacker can craft an attack vector so that a malicious sample can be indistinguishable from an honest sample.
Thus, we develop an GNN model by utilizing IIR GFs to  fit abrupt changes in the spectral domain of the graph structural data.
We utilize existing data-driven techniques in the literature for cyberattack localization to compare our results and tune their hyperparameters using Bayesian optimization technique for a fair comparison.

The contributions of the paper are summarized next:
(i) We utilize IIR GFs in smart grids to efficiently capture the spatial correlations of the graph structural data in a non-Euclidean space for cyberattack localization. The proposed model efficiently predicts the presence of the attack at the bus level.
(ii) We assess by empirical frequency responses of the GFs on IEEE 300-bus test systems; compared to FIR GFs, IIR GFs better approximate the desired filter for the same order and require lower order filters for the same level of approximation.
(iii) We evaluate the localization results both sample-wise and bus-wise by adequately assessing the model performance under various cyberattacks.
E.g., sample wise localization could yield fairly high accuracy for the entire system, yet, the same set of buses could be missed or falsely detected for each sample. Thus, the localization results should be evaluated both sample wise and bus (label) wise to reveal the possible weaknesses.
The rest of this paper is organized as follows.
Section~\ref{problem} presents the preliminaries.
Proposed approach for cyberattack localization is given in Section~\ref{methods}.
Section~\ref{details} presents the numerical experiments. 
Finally, Section~\ref{conclusion} concludes the paper.

\section{Preliminaries}\label{problem} 
The PSSE iteratively solves the following minimization:
\begin{equation} \label{eq:psse}
\hat{\bm{x}} = \argmin\limits_x || \bm{z} - h(\bm{x})||_2^2
\end{equation}
to estimate the system state $\bm{x}$ (voltage and magnitude of each bus) by using complex power measurements $\bm{z}$ (active/reactive bus power injections $\bm{P_i}$, $\bm{Q_i}$ at each bus $i$ and active/reactive branch power flows $\bm{Q_{ij}}$, $\bm{P_{ij}}$ between bus $i$ and $j$) where $h$ represents a nonlinear measurement function which relates $\bm{x}$ to $\bm{z}$.
In FDIA, if an intruder attacks the measurements $\bm{z}$ and can craft an attack vector $\bm{a} = \bm{z_a} - \bm{z} = h(\bm{x+c}) - h(\bm{x})$, then s/he can alter the system state from $\bm{x}$ to $\bm{x+c}$ and inject his/her false data into the system. 

\begin{figure}[h]
	\centering
	\includegraphics[width=0.30\textwidth]{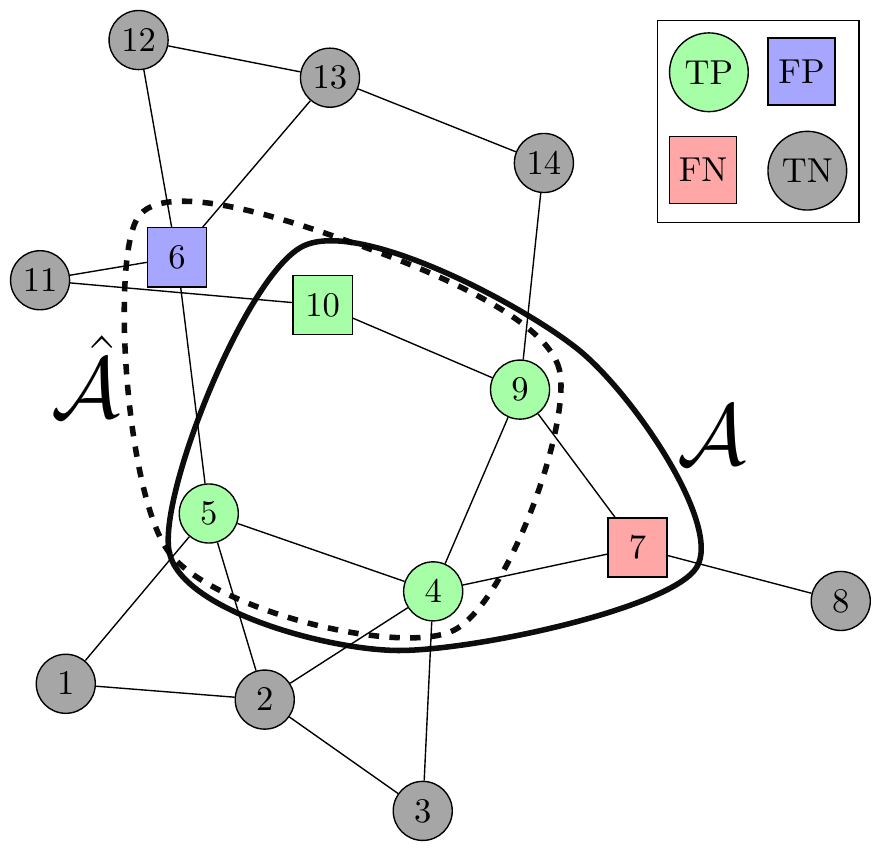}
	\caption{An example attack $\T=\{4,5,7,9,10\}$ and its prediction $\hat{\T}=\{4,5,6,9,10\}$ on the IEEE-14 bus system. In this example, bus 6 is falsely alarmed (false positive) and attack to the bus 7 is missed (false negative).}
	\label{fig:problem}
\end{figure}

In FDIA, an intruder targets a particular region of the grid represented by $\T$ and design the attack vector $\bm{a}$ by altering $\bm{z}$ to spoil $\bm{x}$ in the targeted region.
In contrast, the grid operator tries to detect these attempts and localize the under attack grid area. 
Thus, the localization of the cyberattack problem can be formulated as a \emph{multi-label} classification task in which each bus is equipped with a binary flag to indicate the attack presence. 
The proposed formulation is visualized in Fig.~\ref{fig:problem} by depicting the actual and predicted targeted buses for an example attack on the IEEE 14-bus test system.

\section{IIR Graph Neural Networks for Cyberattack Localization}\label{methods} 
Weighted, undirected, and connected graph $\G = (\V, \E, \bm{W})$ is used to represent the topology of a smart grid, where 
buses are mapped to vertices $\V$, branches and transformers are denoted by edges $\E$, and line admittances are captured by weighted adjacency matrix $\bm{W}$. 
Thus, a signal or function $f$ in $\G$ is represented by a vector $\bm{f} \in \R^n$, where the element $i$ of the vector corresponds to a scalar at the vertex $i \in \V$.

\subsection{Spectral Graph Filters}
Normalized Laplacian
$\bm{L} = \bm{I_n} - \bm{D}^{-1/2} \bm{W} \bm{D}^{-1/2} = \bm{U} \bm{\Lambda} \bm{U}^\top \in \R^{n \times n}$
is a fundamental operator in spectral graph theory. Matrices $\bm{D} \in \R^{n \times n}$,
$\bm{\Lambda} = \diag([\lambda_1, \ldots, \lambda_{n}]) \in \R^{n \times n}$ and $\bm{U} = [\bm{u}_1, \ldots, \bm{u}_{n}] \in \R^{n \times n}$ captures the degrees,  the $n$ eigenvalues associated with the graph Fourier frequencies and the $n$ orthonormal eigenvectors $\bm{u_i}$ representing the graph Fourier basis \cite{ortega2018graph}, respectively.
A vertex (spectral) domain signal is transformed into the spectral (vertex) domain by the forward (inverse) Graph Fourier Transform defined by $\bm{\tilde{x}} = \bm{U}^\top \bm{x}$ ($\bm{x} = \bm{U} \bm{\tilde{x}}$) where $\bm{x}$ ($\bm{\tilde{x}}) \in \R^{n}$ denote the vertex (spectral) domain signal \cite{ortega2018graph}. 
A GF $h$ is convolved with $\bm{x}$:
\begin{equation} \label{eq:gsp}
\bm{Y} = h \ast \bm{X} = h(\bm{L}) \bm{X} = \bm{U} h(\bm{\Lambda}) \bm{U}^\top \bm{X}
\end{equation}
by first transforming $\bm{x}$ into the spectral domain using the forward GFT, then multiplying the Fourier components with $h(\bm{\Lambda}) = \diag[h(\lambda_1), \ldots, h(\lambda_{n})]$, and finally inverting it back to the vertex domain by the inverse GFT \cite{ortega2018graph}.
However, since each $\lambda_i$ is processed for each node, spectral filters are not spatially localized.
In addition, due to eigenvalue decomposition (EVD) of $\bm{L}$ and the matrix multiplications with $\bm{U}$ and $\bm{U}^\top$, they are computationally complex.

\subsection{FIR Graph Filters}
To localize and reduce the spectral GFs' complexity, FIR graph filters were proposed in \cite{defferrard2016convolutional}.
FIR GFs are $K$-localized and their spectral responses assume the form $h_{FIR}(\lambda) = \sum_{k=0}^{K-1} a_k \lambda^k$, where only $K$-hop neighbors of a vertex $v$ are considered to calculate the filter response at $v \in \V$.
Nevertheless, FIR GFs require high-degree polynomials to capture the graph's global structure, and due to the poor interpolation and extrapolation capabilities of high degree polynomials, their ability to capture sharp transitions in the frequency response is limited \cite{loukas2015distributed}.

\subsection{IIR Graph Filters}
\begin{figure}[ht]
	\centering
	\includegraphics[width=0.97\linewidth]{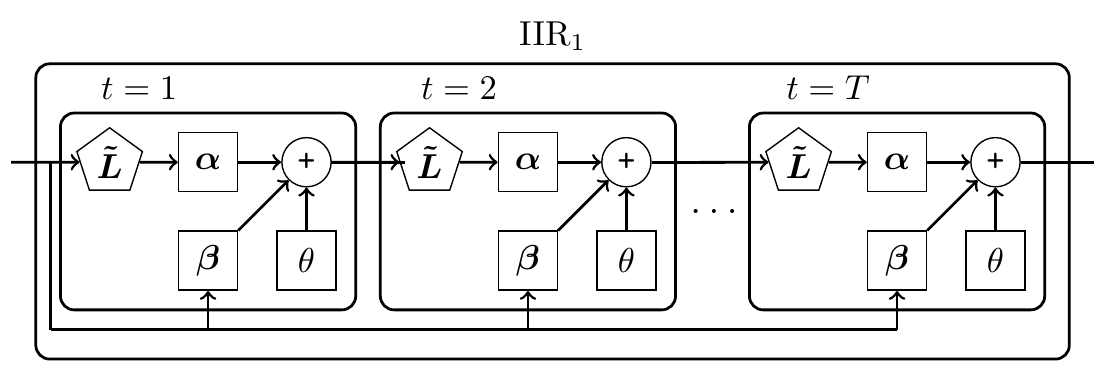}
	\caption{NN implementation of IIR$_1$ GF as a building block of IIR$_K$ GF. In $T$ fixed iterations, an IIR$_1$ block realizes eq.~(\ref{eq:y_arma_2}).}
	\label{fig:arma1}
\end{figure}

To circumvent this limitation, researchers in \cite{loukas2015distributed, shi2015infinite} proposed IIR GFs.
A potential building block of a $K$-order IIR GF is the first-order recursive filter:
\begin{equation}\label{eq:y_arma_1}   
\bm{Y}^{t+1} = a \bm{\tilde{L}} \bm{Y}^t + b \bm{X},
\end{equation}
where $\bm{X}$ denotes the input, $\bm{Y}^t$ is the output at iteration $t$, $a$ and $b$ are arbitrary coefficients, and $\bm{\tilde{L}} = \frac{\lambda_{max} - \lambda_{min}}{2} \bm{I_n} - \bm{L}$ represents the modified Laplacian.
According to  Theorem 1 in \cite{loukas2013think}, eq. (\ref{eq:y_arma_1}) converges regardless of $\bm{Y}^0$ and $\bm{L}$  and its frequency response is given by $h_{ARMA_1}(\tilde{\lambda}_n) = \frac{b}{1 - a \tilde{\lambda}_n}$. 
It can be implemented as a NN layer if we unroll the recursion into $T$ fixed iterations:
\begin{equation}\label{eq:y_arma_2}
\bm{Y}^{t+1} = \bm{\tilde{L}} \bm{Y}^t \bm{\alpha} + \bm{X} \bm{\beta} + \bm{\theta},
\end{equation}
where $\bm{\alpha} \in \R^{c_{out} \times c_{out}}$, $\bm{\beta} \in \R^{c_{in} \times c_{out}}$, and $\bm{\theta} \in \R^{c_{out}}$ are trainable weights, and $c_{in}$ and $c_{out}$ denote the number of channels in the input and  output tensors, respectively.
Since $0 \leq \lambda_{min} \leq \lambda_{max} \leq 2$, the  modified Laplacian can be simplified to $\bm{\tilde{L}} = \bm{I_n} - \bm{L}$ for $\lambda_{min}=0$, and $\lambda_{max}=2$ \cite{bianchi2021graph}.
NN realization of the IIR$_1$ block implementing eq.~(\ref{eq:y_arma_2}) in $T$ fixed iterations is depicted in Fig.~\ref{fig:arma1}.
IIR$_K$ GFs can be implemented by averaging K parallel IIR$_1$ filters: $\bm{Y} = \frac{1}{K}\sum_{k=1}^{K} \bm{Y}_K^\top$, which leads to a rational frequency response  $h_{IIR_{K}}(\tilde{\lambda}_n) = \sum_{k=1}^{K} \frac{b_k}{1 - a_k \tilde{\lambda}_n}$,  with  $K-1$ and $K$ order polynomials at its numerator and denominator, respectively.
Motivated readers are directed to ~\cite{loukas2015distributed, shi2015infinite, loukas2013think, isufi2016autoregressive, bianchi2021graph} for detailed analysis.

\subsection{Architecture of the Proposed Model}
\begin{figure}[h]
	\centering
	\newcommand{\wdt}{0.97} 
	\includegraphics[width=\wdt\linewidth]{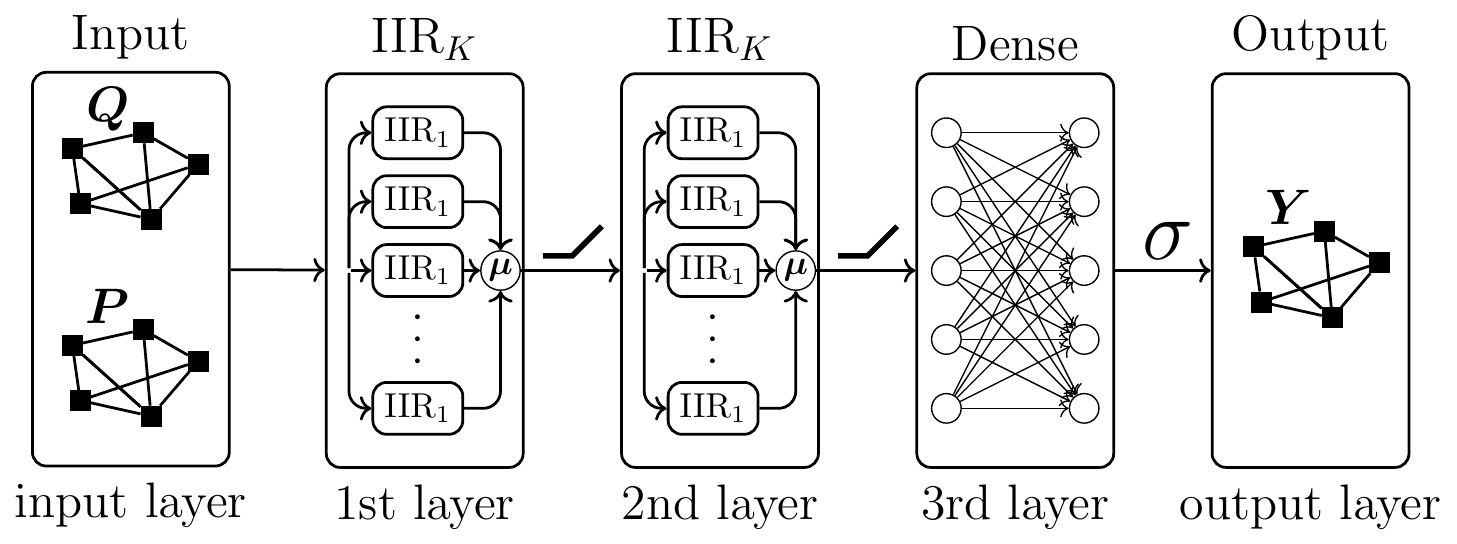}
	\caption{Architecture of the proposed model with three hidden layers where each IIR$_K$ layer consists of $K$ parallel IIR$_1$.}
	\label{fig:arhitecture}
\end{figure}

The proposed model consists of
one input layer to represent complex bus power injections,
$L-1$ hidden IIR$_K$ layers to extract spatial features,
one dense layer to distribute the node features, 
and one output layer to predict the attack at each node.
Fig.~\ref{fig:arhitecture} illustrates the proposed model's architecture for $L=3$ with a small graph having $n=5$.

In the multi-layered architecture,
$\bm{X^0} \in \R^{n \times 2}$ represents the input tensor $[\bm{P}, \bm{Q}]$, 
$\bm{X^l} \in \R^{n \times c_l}$ defines hidden layer $l$'s output tensor,
$\bm{Y} \in \R^{n}$ denotes the model outputs as the location of the attack,
and  $c_l$ identifies the number of channels in layer $l$ for $1 \leq l \leq L$.
Specifically, an IIR$_K$ layer takes $\bm{X}^{l-1} \in \R^{n \times c_{l-1}}$ as its input and produces $\bm{X}^{l} \in \R^{n \times c_{l}}$ as its output in layer $l$,
the dense layer propagates the information to the whole graph and 
the output layer returns the probability of the attack at the node level via $\bm{Y} \in \R^{n}$.
While the ReLU activation is used at the end of each IIR$_K$ layer to increase the model's nonlinear modeling ability, the sigmoid is employed to transform the outputs into  probabilities.

\section{Numerical Experiments}\label{details} 

\subsection{Frequency Response of FIR and  IIR GFs}
To demonstrate that IIR GFs better fit the sharp changes in frequency response compared to FIR GFs,
we design an ideal highpass GF $h^\dagger$ for IEEE 300-bus test system:
\begin{equation} \label{eq:dagger}
{h^\dagger}(\lambda) = 
\begin{cases}
1, & \lambda  > \frac{\lambda_{max}}{2} \\
0, & \text{otherwise}.
\end{cases}
\end{equation}
Note that similar results can be obtained by any other filter or test cases \cite{shi2015infinite}.
Let $\bm{x}, \bm{y} \in R^{n}$ denote the input and output of a GF $h(\lambda)$, respectively.
The empirical frequency response $\tilde{h}$ can be expressed as $\tilde{h}(\lambda_{i}) = \frac{\bm{u}_{i}^\top \bm{y}}{ \bm{u}_{i}^\top \bm{x}}$ \cite{loukas2015distributed}.
Each $\tilde{h}(\lambda_{i})$ shows how $\bm{u}_{i}$, corresponding to $\lambda_{i}$, ``scales'' $\bm{x}$ to obtain $\bm{y}$.

\begin{figure}[h] 
	\centering
	\newcommand{\wdt}{0.72} 
	\subfloat[an ideal highpass $h^\dagger$ for IEEE-300. \label{fig:118_filter}]{
		\centering
		\includegraphics[width=\wdt\linewidth]{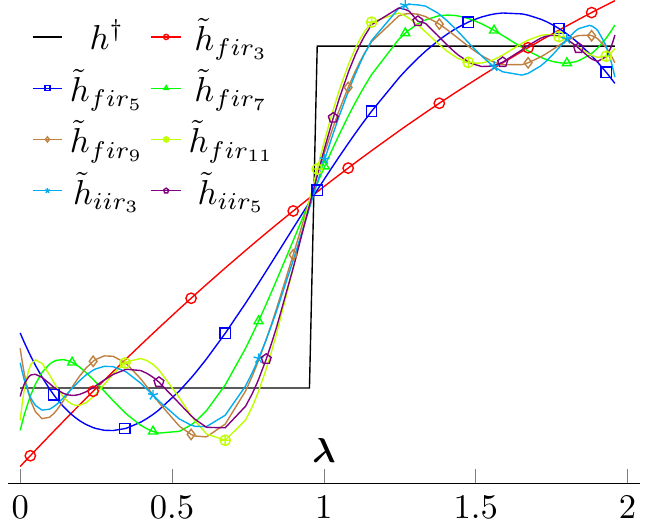}
	}
	\caption{Empirical frequency response of FIR and IIR GFs when approximating an ideal filter $h^\dagger$ applied on IEEE 300-bus test system. Compared to FIR GF, IIR GF better approximates the desired filter for the same $K$ (e.g., $\tilde{h}_{cheb_3}$ vs $\tilde{h}_{arma_3}$) and it requires a lower $K$ for the same level of approximation (e.g., $\tilde{h}_{cheb_{11}}$ vs $\tilde{h}_{arma_5}$).}
	\label{fig:filters}
\end{figure}

As a first step, we randomly generate $2^{16}$ $\bm{x}$s for the aforementioned system from the normal distribution and filter them by $h^\dagger$ using eq. (\ref{eq:gsp}) to obtain $\bm{y}$s.
Next, we train a layer of FIR and IIR GNN models in mini batches with $2^6$ samples of $\bm{x}$ and $\bm{y}$ as the input and output values of the models until there is no further improvement.
Then, we calculate and plot $\tilde{h}(\lambda_{i})$ values for each $\bm{x}, \bm{y}$ tuple \cite{boyaci2021joint}.
Fig.~\ref{fig:filters} demonstrates that IIR GFs are more flexible to fit sudden changes for a fixed $K$ compared to FIR GFs.
It is the main motivation of this paper to select IIR GFs for localizing cyberattacks in smart grids.

\subsection{Data Generation}
Since there is no publicly available dataset in the task of cyberattack localization, we generate a synthetic dataset using historical load profiles. 
First, we download 5-minute intervals of the actual load profile of NYISO for July 2021 and interpolate them to increase the resolution to 1-minute.
Next, we generate a realistic dataset for IEEE 300-bus test system using 1-minute interval load profile. Specifically, the load values are distributed and scaled  proportional to their initial values, AC power flow algorithms are executed, and 1\% noisy power measurements are saved for each timestamp.

To simulate the cyberattacks, we implement some of the frequently used FDIA generation algorithms in the literature, such as data replay attacks ($A_r$) \cite{chaojun2015detecting}, data scale attacks ($A_s$) \cite{jevtic2018physics}, distribution-based ($A_d$) attacks \cite{ozay2015machine}, and optimization based attacks ($A_o$) \cite{boyaci2021graph, boyaci2022cyberattack}.
A measurement $z_o^i$ is replaced with one of its previous values in $A_r$,  
it is multiplied with a number sampled from a uniform distribution ($\U$) between 0.9 and 1.1 in $A_s$, and
it is changed with a value drawn from the Gaussian distribution satisfying the same mean and variance with it in $A_d$.
 $A_o$ solves a constrained optimization problem to maximize the state variables' deviation while minimizing the attack power on measurements.

We shuffle the whole data for seasonality elimination, scale it with the normal scaler for faster training, and split it into three sections: 4/6 for training, 1/6 for validation, and 1/6 for testing the proposed models.
We arbitrarily select $A_o$ and $A_d$ and include them in the training and validation splits to evaluate the performance of our method under unseen attack types.
Test split, on the contrary, includes all four cyberattacks.
To have a balanced classification problem, the number of attacked samples are kept equal with the number of unattacked samples in each split. 
The final dataset has 34560 samples, where each sample consist of complex power measurements and $n$ binary labels for the attack presence at each bus. 

\subsection{Features, Training, and Metrics}
Since $\bm{P_i}+j\bm{Q_i} = \sum_{k \in \Omega_i} \bm{P_{ik}} +j\bm{Q_{ik}}$, node features can represent branch features as summation in their corresponding set of buses $\Omega_i$ connected to bus $i$.
Therefore, we only feed  $\bm{P}$ and $\bm{Q}$ values to the model as seen from the input layer of Fig.~\ref{fig:arhitecture}.
In addition, we select $\bm{W} = |\bm{Y_{bus}}|$ to calculate $\bm{\tilde{L}}$ and feed the IIR$_K$ layers where $\bm{Y_{bus}} \in \R^{n \times n}$ represents the nodal admittance matrix of the power grid.

We use $F1$ score $F1 = \frac{2*TP}{2TP + FP + FN}$ to evaluate model performances, where $TP$, $FP$, $TN$, and $FN$ represent true positives, false positives, true negatives, and false negatives, respectively.
In addition, to overcome the division by zero problem  we assume $F1=1$ when there is no attack at all and all labels are correctly predicted as not attacked.
Otherwise, we assign $F1=0$ even if there is one mismatch. 

We employ multilabel supervised training using the binary cross-entropy loss to compute unknown trainable parameters of the model. 
Training samples are fed into the model as mini batches of 256 samples with 256 maximum number of epochs.
Moreover, we utilize early stopping criteria where 16 epochs are tolerated without any improvement in the validation set's cross entropy loss.
Implementations are carried out in Python 3.8 on Intel i9-8950 HK CPU \@ 2.90GHz with NVIDIA GeForce RTX 2070 GPU.

\subsection{Localization Results} \label{sec:result}
We implement other  data-driven methods from the literature to compare them with our method. To the best of our knowledge, \cite{jevtic2018physics} is the only data-driven approach in the literature in which LSTM based model is used for cyberattack localization. Thus, we train an LSTM based localizer with our dataset to compare model performances. Moreover, despite the fact that they are proposed for cyberattak detection, we implement other  data-driven methods from the literature that are suitable for the multi-label classification task such as Fully Connected Network (FCN) \cite{s96}, Convolutional Neural Network (CNN) \cite{s99}, and FIR-GNN \cite{boyaci2021graph, boyaci2022cyberattack}. We train, validate and test these models similarly to the proposed IIR-GNN model using the generated dataset.
Models are trained on the training set and their hyper-parameters are optimized on the validation set by Bayesian optimization techniques for each model in 250 trials.

Cyberattack localization is a multi-label classification problem, therefore, it can be evaluated in two possible ways: (i) bus-wise (BW) evaluation where each bus is evaluated separately along the samples, and (ii) sample-wise (SW) evaluation where each sample at a fixed time-step is treated individually along the buse. 
Thus, for detailed assessment, we analyze the distributions of BW and SW localization results in $F1$ percentages by the ratio of items satisfying some predetermined thresholds which provides quantifiable metrics to assess model performance.
For instance, the percentage of samples (buses) having $F1 \ge 90\%$ in SW (BW) evaluation are used to measure the ratio of ``acceptable'' samples (buses) in the distributions.

\begin{figure}[h] 
	\centering
	\newcommand{\wdt}{0.97} 
	\includegraphics[width=\wdt\linewidth]{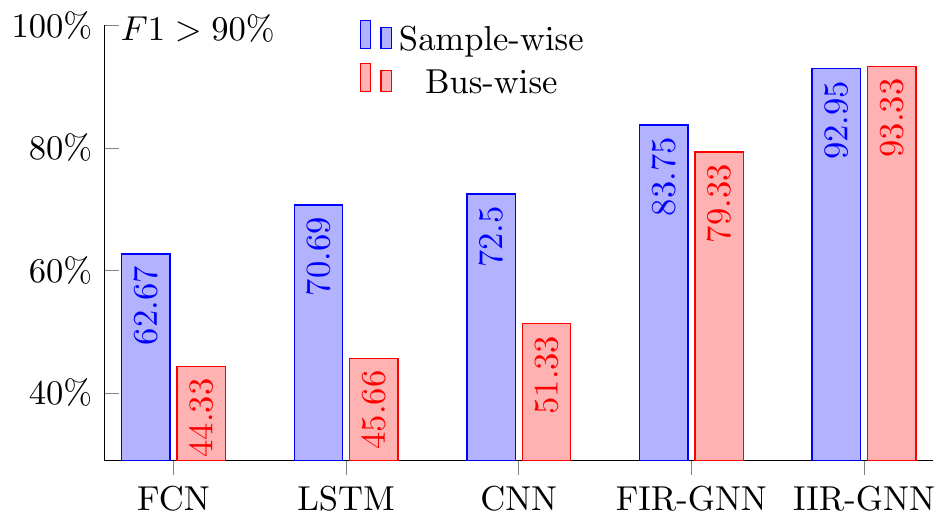}
	\caption{Ratio of $F1$ scores greater than 90\% for sample-wise and bus-wise evaluation of localization.}
	\label{fig:localization}
\end{figure}

Localization results are plotted in Fig.~\ref{fig:localization}.
Only IIR-GNN model reaches $93\%$ $F1$ level in both SW and BW evaluation.
To be specific, in 92.95\% of the samples, the localization ratio is greater than $90\%$ in the IIR-GNN model.
Similarly, in 93.33\% of the buses, which corresponds to 279 buses for IEEE 300-bus test system, attack localization success is greater than $90\%$ $F1$ level.
Its ``acceptable'' ($F1 \ge 90\%$) percentages are $9.2\%$, and $14\%$ greater than the second best model FIR-GNN in SW and BW localization, respectively. 

For the target test system having 300 buses, the localization delays are measured as 1.50, 99.78, 2.73, 2.71, 2.94 milliseconds for the FCN, LSTM, CNN, FIR-GNN, and IIR-GNN models, respectively. So, except for the LSTM architecture, all models can be considered real-time compatible.
\begin{table}[h] 
	\centering
	\caption{Joint detection and localization times in milliseconds.}
	\setlength{\tabcolsep}{3pt}
	\renewcommand{\arraystretch}{1.2}
	\newcolumntype{?}[1]{!{\vrule width #1}}
	\begin{tabular}{ c| c | c | c | c}
		\textbf{FCN} & \textbf{LSTM} & \textbf{CNN} & \textbf{FIR-GNN} & \textbf{IIR-GNN} \\ \specialrule{1pt}{1pt}{1pt}
		1.50 & 99.78 & 2.73 & 2.71 & 2.94 \\
	\end{tabular}
	\label{tab:time}
\end{table}

\section{Conclusion}\label{conclusion}
To effectively model the smart grid data's implicit graph structure, we utilized IIR GNN for cyberattack localization purposes.
As a first step, by comparing their empirical frequency responses we verified that IIR GFs better approximate the desired filter response compared to the FIR GFs.  
Next, we present a multilayered IIR GNN model to localize the cyberattack at the bus level resolution.
Then, we validate the proposed model along with the existing architectures under distinct cyberattack models using both sample-wise (SW) and bus-wise (BW) evaluations.
It is numerically shown that the proposed model surpasses the state-of-the-art FIR GNN model by $9.2\%$ and $14\%$ in SW and BW localization, respectively.

\bibliographystyle{IEEEbib}
\bibliography{eusipco2022}

\begin{thebibliography}{10}

\bibitem{wu2020comprehensive}
Zonghan Wu, Shirui Pan, Fengwen Chen, Guodong Long, Chengqi Zhang, and S~Yu
  Philip,
\newblock ``A comprehensive survey on graph neural networks,''
\newblock {\em IEEE transactions on neural networks and learning systems}, vol.
  32, no. 1, pp. 4--24, 2020.

\bibitem{ortega2018graph}
Antonio Ortega, Pascal Frossard, Jelena Kova{\v{c}}evi{\'c}, Jos{\'e}~MF Moura,
  and Pierre Vandergheynst,
\newblock ``Graph signal processing: Overview, challenges, and applications,''
\newblock {\em Proceedings of the IEEE}, vol. 106, no. 5, pp. 808--828, 2018.

\bibitem{shuman2013emerging}
David~I Shuman, Sunil~K Narang, Pascal Frossard, Antonio Ortega, and Pierre
  Vandergheynst,
\newblock ``The emerging field of signal processing on graphs: Extending
  high-dimensional data analysis to networks and other irregular domains,''
\newblock {\em IEEE signal processing magazine}, vol. 30, no. 3, pp. 83--98,
  2013.

\bibitem{defferrard2016convolutional}
Michaël Defferrard, Xavier Bresson, and Pierre Vandergheynst,
\newblock ``Convolutional neural networks on graphs with fast localized
  spectral filtering,''
\newblock in {\em Advances in Neural Information Processing Systems (NIPS)},
  2016.

\bibitem{bianchi2021graph}
Filippo~Maria Bianchi, Daniele Grattarola, Lorenzo Livi, and Cesare Alippi,
\newblock ``Graph neural networks with convolutional arma filters,''
\newblock {\em IEEE Transactions on Pattern Analysis and Machine Intelligence},
  2021.

\bibitem{shi2015infinite}
Xuesong Shi, Hui Feng, Muyuan Zhai, Tao Yang, and Bo~Hu,
\newblock ``Infinite impulse response graph filters in wireless sensor
  networks,''
\newblock {\em IEEE Signal Processing Letters}, vol. 22, no. 8, pp. 1113--1117,
  2015.

\bibitem{yu2016smart}
Xinghuo Yu and Yusheng Xue,
\newblock ``Smart grids: A cyber--physical systems perspective,''
\newblock {\em Proceedings of the IEEE}, vol. 104, no. 5, pp. 1058--1070, 2016.

\bibitem{abur2004power}
A.~Abur and A.G. Exp{\'o}sito,
\newblock {\em Power System State Estimation: Theory and Implementation},
\newblock Power Engineering (Willis). CRC Press, 2004.

\bibitem{musleh2019survey}
Ahmed~S Musleh, Guo Chen, and Zhao~Yang Dong,
\newblock ``A survey on the detection algorithms for false data injection
  attacks in smart grids,''
\newblock {\em IEEE Transactions on Smart Grid}, vol. 11, no. 3, pp.
  2218--2234, 2019.

\bibitem{nudell2015real}
Thomas~R Nudell, Seyedbehzad Nabavi, and Aranya Chakrabortty,
\newblock ``A real-time attack localization algorithm for large power system
  networks using graph-theoretic techniques,''
\newblock {\em IEEE Transactions on Smart Grid}, vol. 6, no. 5, pp. 2551--2559,
  2015.

\bibitem{khalaf2018joint}
Mohsen Khalaf, Amr Youssef, and Ehab El-Saadany,
\newblock ``Joint detection and mitigation of false data injection attacks in
  agc systems,''
\newblock {\em IEEE Transactions on Smart Grid}, vol. 10, no. 5, pp.
  4985--4995, 2018.

\bibitem{hasnat2020detection}
Md~Abul Hasnat and Mahshid Rahnamay-Naeini,
\newblock ``Detection and locating cyber and physical stresses in smart grids
  using graph signal processing,''
\newblock {\em arXiv preprint arXiv:2006.06095}, 2020.

\bibitem{jevtic2018physics}
Ana Jevtic, Fengli Zhang, Qinghua Li, and Marija Ilic,
\newblock ``Physics-and learning-based detection and localization of false data
  injections in automatic generation control,''
\newblock {\em IFAC-PapersOnLine}, vol. 51, no. 28, pp. 702--707, 2018.

\bibitem{liu2011false}
Yao Liu, Peng Ning, and Michael~K Reiter,
\newblock ``False data injection attacks against state estimation in electric
  power grids,''
\newblock {\em ACM Transactions on Information and System Security (TISSEC)},
  vol. 14, no. 1, pp. 1--33, 2011.

\bibitem{loukas2015distributed}
Andreas Loukas, Andrea Simonetto, and Geert Leus,
\newblock ``Distributed autoregressive moving average graph filters,''
\newblock {\em IEEE Signal Processing Letters}, vol. 22, no. 11, pp.
  1931--1935, 2015.

\bibitem{loukas2013think}
Andreas Loukas, Marco Zuniga, Matthias Woehrle, Marco Cattani, and Koen
  Langendoen,
\newblock ``Think globally, act locally: On the reshaping of information
  landscapes,''
\newblock in {\em Proceedings of the 12th international conference on
  Information processing in sensor networks}, 2013, pp. 265--276.

\bibitem{isufi2016autoregressive}
Elvin Isufi, Andreas Loukas, Andrea Simonetto, and Geert Leus,
\newblock ``Autoregressive moving average graph filtering,''
\newblock {\em IEEE Transactions on Signal Processing}, vol. 65, no. 2, pp.
  274--288, 2016.

\bibitem{boyaci2021joint}
Osman Boyaci, Mohammad~Rasoul Narimani, Katherine Davis, Muhammad Ismail,
  Thomas~J Overbye, and Erchin Serpedin,
\newblock ``Joint detection and localization of stealth false data injection
  attacks in smart grids using graph neural networks,''
\newblock {\em IEEE Transactions on Smart Grid}, pp. 1--1, 2021.

\bibitem{chaojun2015detecting}
Gu~Chaojun, Panida Jirutitijaroen, and Mehul Motani,
\newblock ``Detecting false data injection attacks in ac state estimation,''
\newblock {\em IEEE Transactions on Smart Grid}, vol. 6, no. 5, pp. 2476--2483,
  2015.

\bibitem{ozay2015machine}
Mete Ozay, Inaki Esnaola, Fatos Tunay~Yarman Vural, Sanjeev~R Kulkarni, and
  H~Vincent Poor,
\newblock ``Machine learning methods for attack detection in the smart grid,''
\newblock {\em IEEE transactions on neural networks and learning systems}, vol.
  27, no. 8, pp. 1773--1786, 2015.

\bibitem{boyaci2021graph}
Osman Boyaci, Amarachi Umunnakwe, Abhijeet Sahu, Mohammad~Rasoul Narimani,
  Muhammad Ismail, Katherine~R. Davis, and Erchin Serpedin,
\newblock ``Graph neural networks based detection of stealth false data
  injection attacks in smart grids,''
\newblock {\em IEEE Systems Journal}, pp. 1--12, 2021.

\bibitem{boyaci2022cyberattack}
Osman Boyaci, M~Rasoul Narimani, Katherine Davis, and Erchin Serpedin,
\newblock ``Cyberattack detection in large-scale smart grids using chebyshev
  graph convolutional networks,''
\newblock in {\em 2022 9th International Conference on Electrical and
  Electronics Engineering (ICEEE)}. IEEE, 2022, pp. 217--221.

\bibitem{s96}
Yi~Wang, Mahmoud~M Amin, Jian Fu, and Heba~B Moussa,
\newblock ``A novel data analytical approach for false data injection
  cyber-physical attack mitigation in smart grids,''
\newblock {\em IEEE Access}, vol. 5, pp. 26022--26033, 2017.

\bibitem{s99}
Defu Wang, Xiaojuan Wang, Yong Zhang, and Lei Jin,
\newblock ``Detection of power grid disturbances and cyber-attacks based on
  machine learning,''
\newblock {\em Journal of Information Security and Applications}, vol. 46, pp.
  42--52, 2019.

\end{thebibliography}

\end{document}